\documentclass{osa-article}
\journal{josaa}

\usepackage{bm}
\usepackage{float}
\usepackage[list=true]{subcaption}
\usepackage{algorithm}
\usepackage[titletoc]{appendix}
\usepackage{xcolor}
\usepackage{cite}
\usepackage{pgfplots}
\usetikzlibrary{intersections}

\DeclareMathOperator*{\argmin}{argmin}

\DeclareMathOperator{\diver}{div}
\newcommand{\tran}{^{\mathsf{T}}}

\newcommand{\diff}{\mathrm{d}}
\newcommand{\Diff}{\mathrm{D}}
\newcommand{\Sc}{\mathcal{S}}
\newcommand{\Tc}{\mathcal{T}}
\newcommand{\Xc}{\mathcal{X}}
\newcommand{\Yc}{\mathcal{Y}}

\newcommand{\sray}{\bm{\hat{s}}}
\newcommand{\tray}{\bm{\hat{t}}}
\newcommand{\nvec}{\bm{\hat{n}}}

\newcommand{\gt}{g_\mathrm{t}}

\begin{document}
\title{Fresnel reflections in inverse freeform lens design}

\author{A.H.~van~Roosmalen,\authormark{1,*} M.J.H.~Anthonissen,\authormark{1} W.L.~IJzerman,\authormark{1,2} J.H.M.~ten~Thije~Boonkkamp\authormark{1}}

\address{\authormark{1}CASA, Department of Mathematics and Computer Science, Eindhoven University of Technology, \\PO Box 513, 5600 MB Eindhoven, The Netherlands\\
\authormark{2}Signify Research, High Tech Campus 7, 5656 AE Eindhoven, The Netherlands\\
\authormark{*}Corresponding author: \email{a.h.v.roosmalen@tue.nl}}

\begin{abstract}
  In this paper we propose a method to design a freeform lens including the effect of Fresnel reflections on the transmitted intensity. This method is elaborated for a lens with one freeform surface shaping a far-field target from a point source or collimated input beam. It combines the optical mapping with the energy balance incorporating the loss due to Fresnel reflections, which leads to a generalized Monge-Amp\`ere equation. We adapt a least-squares solver from previous research to solve the model numerically. This is then tested with a theoretical example and a test case related to road lighting.
\end{abstract}

\section{Introduction}
Recently, a lot of research has been done on inverse methods for freeform optical design. Especially freeform lenses have become more popular with the introduction of LED lighting. This has a lower working temperature and allows for the usage of plastic materials, which are easier to manufacture in arbitrary shapes. The arbitrary shapes of freeform lenses give rise to arbitrary angles of incidence at the freeform surface. The fraction of reflected light can strongly vary with these angles \cite[Sec.~4.6]{hecht16}. Especially close to the critical angle, a large part of the incident flux is deflected into this so-called Fresnel reflection. Most inverse methods for freeform design do not take into account these Fresnel reflections, while they can have a significant influence on the outgoing intensity distribution. In this paper we will introduce a method that does take into account Fresnel reflections. \\
\indent There have been many methods developed for designing freeform optics recently. We give a brief overview of some of these methods. For a more complete summary, see \cite{rom19}. Some approaches use numerical methods such as finite differences and Newton's method to solve the Monge-Amp\`ere equation related to the optimal transport formulation of the problem \cite{bos17,bos18,wu18}. The supporting quadrics method was proposed by Oliker et al. \cite{oli15} to design freeform surfaces. An alternative approach based on discretizing the corresponding optimal transport problem into a linear assignment problem has been developed by Doskolovich et al. \cite{dosko19}. A ray mapping method has also been used to solve this optical design problem for arbitrary wavefronts \cite{feng19,wei19}. \\
\indent As mentioned, most inverse methods for freeform optical design do not take into account Fresnel reflections. There are some methods that aim to minimize the total loss due to Fresnel effects \cite{wei19b,shen21,gan18}. However, the authors only try to minimize the reflectance, and do not take into account the effect of the Fresnel reflections on the outgoing intensity or irradiance. To the best of our knowledge, only one paper claims to incorporate this effect in their freeform design method, but it is not explained how this is done \cite{bad21}. \\
\indent A least-squares algorithm has been developed by Prins et al. as a way to design a freeform surface for transforming a collimated beam into a far-field target \cite{prins15,thije19}. This has later been expanded to shape one collimated beam to another \cite{yad19}, to create a far-field target from a point source \cite{rom19,rom20} and to collimate and shape a beam from a point source \cite{teun21}. A more complete overview of optical systems for which these methods have been adapted has been given by Romijn \cite{rom21} and by Anthonissen et al. \cite{ant21}. In this paper we will adapt the least-squares algorithm as in \cite{rom19} to account for Fresnel losses. These changes are made to give an actual transmitted intensity of the same shape as a given hypothetical target distribution without Fresnel reflection, up to scaling, despite variations in the reflectance. The loss is unknown beforehand, so the scaling factor is as well. We apply our adapted algorithm to two different lens systems. Both have a far-field target, one having a point source and the other having a collimated input beam. \\
\indent In this work, we first derive the equations for the relation between source distribution, target distribution and the freeform lens surface in Section \ref{system}. We also derive a convenient expression for the reflectance. In Section \ref{ls} we present an algorithm to solve the aforementioned equations for the surface shape. Since parts of the algorithm remain unchanged, we give a brief summary and refer to other papers for more details. In Section \ref{results} we show the results of our algorithm on two test cases. One theoretical example consists of a uniform source and target, while the other has a practical application as road lighting.

\section{The optical systems}\label{system}
In this section we will first give a mathematical description of the optical systems. Then, we will derive the equations necessary for computing the optical surfaces. This includes the reflection coefficients, which we will write in a form that is convenient to include in our algorithm.
\subsection{Optical system layouts}
We consider two optical systems consisting of a lens, shaping the light distribution from a point source or parallel beam to a far-field target. Both cases are sketched in Figure \ref{fig:sketch}. The lens, with refractive index $n$, has a surface perpendicular to the incident rays and a freeform surface where the light exits the lens. In other words, only the second surface shapes the output distribution. We assume that the surrounding medium has refractive index $1$. The unit direction vectors of a ray before and after the freeform surface are given by $\sray=(s_1,s_2,s_3)\tran$ and $\tray=(t_1,t_2,t_3)\tran$, respectively.
\begin{figure}[ht!]
\begin{tikzpicture}

\def\X{2};
\def\Y{1.5};
\def\R{0.7};

\filldraw [black] (0,0) circle (2pt)
    node[above left=-1] {$\mathcal{O}$};
\filldraw[fill = blue!20, draw = blue] (\R,0) arc (0:180:\R) -- (-\X,0) node[pos=0.5,above]{$n$} arc (180:0:{\X} and \Y) -- cycle;
\draw[blue,thick] (-\X,0) arc (180:0:{\X} and \Y)
    node[pos=0.3,align=center,above left=0] {Freeform surface\\$\bm{r}(\sray)=u(\sray)\sray$};

\foreach \x in {-4,...,1}
    \draw[black] (\x,0) -- ++(-50:1);
\def\lsub{0.5/cos(50)};
\draw[black] (2,0) -- ++(-50:{\lsub});
\draw[black,semithick] (-4,0)--(2.5,0)
    node[fill=white,near start,below=4] {Substrate};

\draw[black,thick,->] (0,0) -- (70:{\X} and \Y) -- ++(40:1) node[right] {$\tray$};
\draw[black,thick,->] (70:{\X} and \Y) -- ++(-70:1);
\draw[black,thick,->,scale = {1/sqrt((cos(70)*\X)^2+(sin(70)*\Y)^2)}] (0,0) -- (70:2 and 1.5) node[left] {$\sray$};

\end{tikzpicture}
\hfill
\begin{tikzpicture}[baseline = -.8cm]
  \draw[thick,->] (0,0) -- (0.5,0) node[below,near end] {$x$};
  \draw[thick,->] (0,0) -- (0,0.5) node[right,near end] {$z$};
\end{tikzpicture}
\hfill
\begin{tikzpicture}

\def\X{2};
\def\Y{1.5};
\def\R{0.7};

\filldraw[fill = blue!20, draw = blue,yshift = 0.5cm] (45:{\X} and \Y) arc (45:135:{\X} and \Y) -- ++(0,-1) node[right,near end] {$n$} -| cycle;
\path[name path=arc1,draw,blue,thick,yshift = 0.5cm] (45:{\X} and \Y) arc (45:135:{\X} and \Y)
    node[pos=0.7,align=center,above left=0] {Freeform surface\\$z=u(\bm{x})$};

\foreach \x in {-3,...,1}
    \draw[black] (\x,0) -- ++(-50:1);
\def\lsub{0.5/cos(50)};
\draw[black] (2,0) -- ++(-50:{\lsub});
\draw[black,semithick] (-3.5,0)--(2.5,0)
    node[fill=white,very near start,below=4] {Substrate};
\draw[black,very thick] ({-cos(45)*\X},0) -- ({cos(45)*\X},0)
    node[fill=white,pos=0.5,below=2] {Source plane};

\path[name path=ray1,draw=none] (0.4,0) -- ++(0,\Y +0.5);
\path [name intersections={of = arc1 and ray1}];
  \coordinate (A)  at (intersection-1);

\filldraw [black] (0.4,0) circle (2pt)
    node[above left] {$\bm{x}$};
\draw[black,thick,->] (0.4,0) -- (A) -- ++(120:1) node[right] {$\tray$};
\draw[black,thick,->] (0.4,0) -- (0.4,1) node[right] {$\sray$};
\draw[black,thick,->] (A) -- ++(240:1);

\draw[very thin] (0.6,0) |- (0.4,0.2);

\end{tikzpicture}
\caption{Projections on the $x,z$-plane of example lenses for a point source (left) and a collimated source beam (right).}
\label{fig:sketch}
\end{figure}
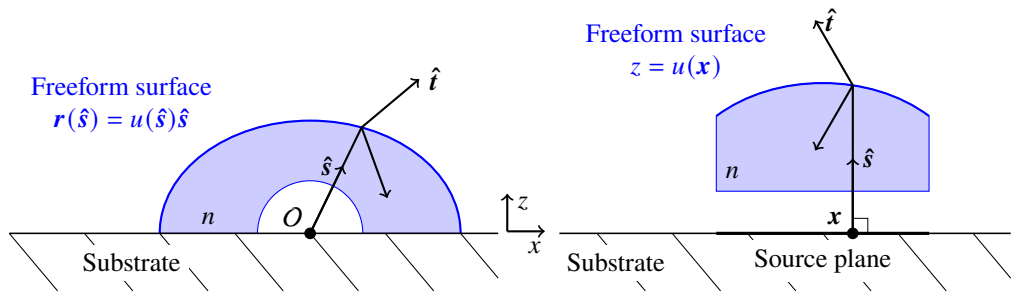\\
\indent For a point source, we have a given source luminous intensity $f=f(\sray)$ and a desired target distribution $g=g(\tray)$, both with a total flux equal to 1. Note that $g$ is a purely hypothetical intensity that can only be achieved if we would disregard Fresnel reflections. The source is located at the origin. The location of the freeform surface is defined by the radial distance from the point source, $u=u(\sray)$. A point on this surface is described by the position vector $\bm{r}(\sray)=u(\sray)\sray$. We introduce the stereographic coordinates from the south pole of the unit sphere $\bm{x}=\bm{x}(\sray)$, so that every $\sray$ except $(0,0,-1)\tran$ has a unique two-dimensional parametrization \cite[Sec.~3.1.2]{rom21}. This transformation is given by
\begin{equation}\label{eq:stereo point}
  \bm{x}(\sray)=\begin{pmatrix}x_1\\x_2\end{pmatrix}=\frac{1}{1+s_3}\begin{pmatrix}s_1\\s_2\end{pmatrix},\qquad\sray(\bm{x})=\frac{1}{1+|\bm{x}|^2}\begin{pmatrix}2x_1\\2x_2\\1-|\bm{x}|^2\end{pmatrix}.
\end{equation}
Analogously, we define the stereographic coordinates $\bm{y}=\bm{y}(\tray)$ for the target. The source domain for vectors $\sray$ is $\Sc\subset S^2$ and contains the support of $f$. The source domain in stereographic coordinates is $\Xc=\bm{x}(\Sc)$. The target domain $\Tc$ is equal to the support of $g$ and its stereographic projection is denoted by $\Yc=\bm{y}(\Tc)$. \\
\indent In the case of a parallel source beam, we have $\sray=(0,0,1)\tran$ for every ray. Instead of the stereographic coordinates, $\bm{x}$ here denotes the position vector on the source plane. The source emittance is given by $f=f(\bm{x})$ and $\Xc$ is the spatial domain on the source plane. We also use a different, generalized stereographic projection for the target coordinates in this system. Instead of a projection from the south pole, we project from the point $(0,0,n)$ above the north pole, followed by a multiplication with $1/n$. This leads to simpler equations later \cite{thije19}. In this case, the generalized stereographic projection is given by
\begin{subequations}\label{eq:stereo par main}
\begin{equation}\label{eq:stereo par}
  \bm{y}(\tray) = \begin{pmatrix} y_1 \\ y_2 \end{pmatrix} = \frac{1}{n-t_3}\begin{pmatrix} t_1 \\ t_2 \end{pmatrix}.
\end{equation}
It can be shown that the inverse is given by
\begin{equation}\label{eq:stereo par inv}
  \tray(\bm{y}) = n\bm{\hat{e}}_z+H(|\bm{y}|;n)\begin{pmatrix}
                                                 y_1 \\
                                                 y_2 \\
                                                 -1
                                               \end{pmatrix}\quad\text{if}\quad (n^2-1)|\bm{y}|^2\leq1,
\end{equation}
where
\begin{equation}
  H(z;n) = \frac{n^2-1}{n+\sqrt{1-(n^2-1)z^2}}.
\end{equation}
\end{subequations}
The condition in Eq. (\ref{eq:stereo par inv}) means that total internal reflection does not happen. \\
\indent When light is incident on an optical surface, a part of it is transmitted while the rest is reflected due to Fresnel reflections. Because the incident rays are perpendicular to the first surface, the reflectance coefficient is constant there. The freeform second surface has a reflectance depending on the incident and transmitted angles \cite[Sec.~4.6]{hecht16}. We assume that all reflected light, from both the inner and freeform surfaces, ends up on the substrate at $z=0$ and is absorbed there.

\subsection{The optical mapping}\label{sec:mapping}
In this section, we state a short summary of the derivation of a relation between the source and target coordinates, implicitly defining a mapping. For a point source, a more in depth derivation is given by Romijn et al. \cite{rom19}. The relation is similar and of the same form for a parallel source \cite{thije19}. \\
\indent Using Hamilton's characteristics \cite{lun64}, it is possible to find the equation
\begin{equation}\label{eq:cost dir}
  \tilde{u}_1(\sray)+\tilde{u}_2(\tray) = \tilde{c}(\sray,\tray).
\end{equation}
The function $\tilde{c}$ is called a cost function and is of the form $\tilde{c}(\sray,\tray)=-\log(n-\sray\bm{\cdot}\tray)$. The functions $\tilde{u}_1$ and $\tilde{u}_2$ are related to the shape of the lens, with $\tilde{u}_1(\sray)=\log u(\sray)$ and $\tilde{u}_2(\tray)$ an auxiliary variable depending on the target coordinates. We change Eq. (\ref{eq:cost dir}) to stereographic coordinates by introducing $u_1(\bm{x})=\tilde{u}_1\big(\sray(\bm{x})\big)$, $u_2(\bm{y})=\tilde{u}_2\big(\tray(\bm{y})\big)$ and $c(\bm{x},\bm{y})=\tilde{c}\big(\sray(\bm{x}),\tray(\bm{y})\big)$ and obtain
\begin{equation}\label{eq:cost}
  u_1(\bm{x})+u_2(\bm{y})=c(\bm{x},\bm{y})=-\log\big(n-\sray(\bm{x})\bm{\cdot}\tray(\bm{y})\big).
\end{equation}
For a parallel source beam we can derive an equation of the same form \cite{thije19}. As mentioned, $\bm{x}$ is then the position coordinate on the source plane and $\bm{y}$ is the generalized stereographic projection defined in Eq. (\ref{eq:stereo par}). In this case, $c(\bm{x},\bm{y})=\bm{x\cdot y}$, $u_1=u$ and $u_2$ is a function containing Hamilton's mixed characteristic.\\
\indent Eq. (\ref{eq:cost}) has many solutions, so we can choose a specific one. A unique solution can be found by assuming that $u_1$ and $u_2$ are a $c$-convex or $c$-concave pair \cite{yad18}. The $c$-convex solution pair has the form
\begin{equation}
  u_1(\bm{x}) = \max_{\bm{y}\in\Yc}\big(c(\bm{x},\bm{y})-u_2(\bm{y})\big),\qquad u_2(\bm{y})=\max_{\bm{x}\in\Xc}\big(c(\bm{x},\bm{y})-u_1(\bm{x})\big).
\end{equation}
Similarly, the $c$-concave solution pair has the form
\begin{equation}
  u_1(\bm{x}) = \min_{\bm{y}\in\Yc}\big(c(\bm{x},\bm{y})-u_2(\bm{y})\big),\qquad u_2(\bm{y})=\min_{\bm{x}\in\Xc}\big(c(\bm{x},\bm{y})-u_1(\bm{x})\big).
\end{equation}
As a result, with either choice, $u_2$ is obtained as a stationary point of $c({\,\cdot\,},\bm{y})-u_1$ with respect to $\bm{x}$, so we have the necessary condition
\begin{equation}\label{eq:nabla}
  \nabla_{\bm{x}}c(\bm{x},\bm{y})-\nabla u_1(\bm{x})=\bm{0}.
\end{equation}
This equation implicitly defines a mapping $\bm{y}=\bm{m}(\bm{x})$, provided that $\bm{C}=\bm{C}(\bm{x},\bm{y})=\Diff_{\bm{xy}}c=\left(\frac{\partial^2 c}{\partial x_i\partial y_j}\right)$ is a regular matrix, by the implicit function theorem \cite[Sec.~12.8]{ada13}. However, the computation of $\bm{m}$ can be very difficult, especially for the system with the point source. Instead of computing the mapping directly, we derive an equation for the Jacobian of $\bm{m}$. This will be useful later, see Sec. \ref{sec:energy}. To achieve this, we substitute $\bm{y}=\bm{m}(\bm{x})$ into Eq. (\ref{eq:nabla}) and take the derivative with respect to $\bm{x}$ again to obtain
\begin{equation}\label{eq:Dm}
  \bm{C}\Diff\bm{m} = \Diff^2u_1-\Diff_{\bm{xx}}c =:\bm{P},
\end{equation}
where $\Diff\bm{m}$ is the Jacobian matrix of $\bm{m}$, $\Diff^2u_1$ is the Hessian matrix of $u_1$ and $\Diff_{\bm{xx}}c=\Diff^2c({\,\cdot\,},\bm{y})$. The Hessian matrix of $c({\,\cdot\,},\bm{y})-u_1$ is equal to $-\bm{P}$, so a $c$-convex pair $u_1$, $u_2$ has a symmetric positive definite (SPD) matrix $\bm{P}$. Similarly, $\bm{P}$ is symmetric negative definite (SND) for a $c$-concave solution pair.

\subsection{Reflection and transmission}
As stated by Hecht \cite[Sec.~4.6]{hecht16}, the reflection coefficient or reflectance $R$ is the fraction of the flux that gets reflected at an optical surface. It depends on the angles of incidence ($\theta_\text{i}$) and refraction ($\theta_\text{t}$) as well as the polarization of the light. We assume here that the light is unpolarized, but the algorithm can easily be adjusted to incorporate different polarizations. The reflectance for light moving from a medium with refractive index $n_\text{i}$ to a medium with index $n_\text{t}$ is then given by
\begin{subequations}\label{eq:refl angles}
\begin{equation}
R = \tfrac12(R_\text{S}+R_\text{P}),
\end{equation}
with $R_\text{S}$ and $R_\text{P}$ the reflectance coefficients for perpendicular and parallel polarized light, respectively, given by
\begin{equation}
  R_\text{S} = \left(\frac{n_\text{i}\cos\theta_\text{i}-n_\text{t}\cos\theta_\text{t}}{n_\text{i}\cos\theta_\text{i}+n_\text{t}\cos\theta_\text{t}}\right)^2,\quad R_\text{P} = \left(\frac{n_\text{t}\cos\theta_\text{i}-n_\text{i}\cos\theta_\text{t}}{n_\text{t}\cos\theta_\text{i}+n_\text{i}\cos\theta_\text{t}}\right)^2.
\end{equation}
\end{subequations}
At the first surface each ray has normal incidence, so the reflection coefficient $R$ is constant there. We denote this coefficient by $R_1$ and using that $n_\text{i}=1$, $n_\text{t}=n$, and $\theta_\text{i}=\theta_\text{t}=0$, we obtain from the above equation
\begin{equation}
  R_1 = \left(\frac{1-n}{1+n}\right)^2.
\end{equation}
At the second surface, rays go from $n_\text{i}=n$ to $n_\text{t}=1$. The rays are generally not normal to the surface. To calculate the reflectance in a point on this surface using Eq. (\ref{eq:refl angles}) we need the angle between $\sray$ and the normal $\nvec$, as well as the angle between $\tray$ and $\nvec$. For this we need to calculate the surface from the mapping $\bm{m}$. Instead, we rewrite the coefficient as a function $R_2=R_2(\sray,\tray)$, where $\tray$ can easily be computed as the (generalized) inverse stereographic projection of $\bm{m}$. We substitute $\cos\theta_\text{i}=-\sray\bm{\cdot}\nvec$ and $\cos\theta_\text{t}=-\tray\bm{\cdot}\nvec$. We then eliminate $\nvec$ from the expression by using that $\nvec$ is parallel to $\tray-n\sray$, due to Snell's law in vector form. We obtain
\begin{equation}\label{eq:refl point}
\begin{aligned}
  R_2(\sray,\tray) &= \frac{1}{2}\left(\frac{(n\sray-\tray)\bm{\cdot}\nvec}{(n\sray+\tray)\bm{\cdot}\nvec}\right)^2 +\frac{1}{2}\left(\frac{(\sray-n\tray)\bm{\cdot}\nvec}{(\sray+n\tray)\bm{\cdot}\nvec}\right)^2 \\
  &= \frac{1}{2}\left(\frac{(n\sray-\tray)\bm{\cdot}(\tray-n\sray)}{(n\sray+\tray)\bm{\cdot}(\tray-n\sray)}\right)^2 +\frac{1}{2}\left(\frac{(\sray-n\tray)\bm{\cdot}(\tray-n\sray)}{(\sray+n\tray)\bm{\cdot}(\tray-n\sray)}\right)^2 \\
  &= \frac{1}{2(1-n^2)^2}\left[\left(2n\sray\bm{\cdot}\tray-(1+n^2)\right)^2+\frac{1}{(\sray\bm{\cdot}\tray)^2}\left((1+n^2)\sray\bm{\cdot}\tray-2n\right)^2\right].
\end{aligned}
\end{equation}
This expression holds for both source types. For the case of a collimated source beam we can simplify it somewhat. All rays are then parallel to the $z$-axis, so $\sray=(0,0,1)\tran$ and $\sray\bm{\cdot}\tray=t_3$. This gives the reflectance
\begin{equation}\label{eq:refl par}
  R_2(\tray) = \frac{1}{2(1-n^2)^2}\left[\left(2nt_3-(1+n^2)\right)^2+\frac{1}{t_3^2}\left((1+n^2)t_3-2n\right)^2\right].
\end{equation}
We can directly relate the transmission coefficient to the reflection coefficient. Since we assume that no light is absorbed by the lens, all light is either reflected or transmitted. We define $T_1=1-R_1$ and $T_2(\sray,\tray)=1-R_2(\sray,\tray)$ as the transmittance at the first and second surface, respectively.

\subsection{Energy conservation}\label{sec:energy}
The main result from Sec. \ref{sec:mapping}, Eq. (\ref{eq:Dm}), describes the propagation of transmitted rays, but not the flux at the source or target. To take that into account, we need to satisfy conservation of energy. First, we look at the point source. Without Fresnel reflections, the flux in any subset of $\Sc$ should be equal to that contained in its image on $\Tc$. With the Fresnel reflections taken into account, only the flux that is transmitted by both surfaces ends up at the target. Let $\mathcal{A}$ be a subset of $\Sc$ and $\tray(\mathcal{A})\subseteq\Tc$ be its image. When we use the far-field approximation, conservation of energy is given by
\begin{equation}\label{eq:energy int}
  \int_{\mathcal{A}}T_1\,T_2(\sray,\tray)\,f(\sray)\,\diff S(\sray) = \int_{\tray(\mathcal{A})}\gt(\tray)\,\diff S(\tray),
\end{equation}
where $\gt$ is the transmitted target distribution. We choose this to be a scaling of $g$. This hypothetical target distribution $g$ has a total flux equal to that of $f$ (i.e. 1). The transmitted flux can never be equal to that due to the reflected light being absorbed by the substrate. Instead, we want a target distribution with the same shape as $g$, but a flux adapted to the transmission. For that, we choose $\gt=\beta g$, with $\beta\in(0,1)$ the fraction of transmitted flux, dependent on $\bm{m}$. Using the equation above with $\mathcal{A}=\Sc$ and using that the total flux of $g$ is 1, we obtain
\begin{equation}\label{eq:beta}
  \beta(\bm{m}) = T_1\int_{\Sc}T_2\big(\sray,\tray(\bm{m})\big)f(\sray)\,\diff S(\sray).
\end{equation}
\indent Next, we want to use Eq. (\ref{eq:energy int}) to find a Monge-Amp\`ere type equation similar to the one found in previous work \cite{rom19}. We apply substitution laws for integration to write both sides as integrals over $\bm{x}$. This gives us for any $X\subset\Xc$
\begin{equation}
  \int_X T_1\,\tilde{T_2}\left(\bm{x},\bm{m}(\bm{x})\right)J_{\sray}(\bm{x})\tilde{f}(\bm{x})\,\diff\bm{x} = \int_X |\det\big(\Diff\bm{m}(\bm{x})\big)| J_{\tray}\big(\bm{m}(\bm{x})\big)\tilde{\gt}\big(\bm{m}(\bm{x})\big)\,\diff\bm{x},
\end{equation}
with $\tilde{f}(\bm{x})=f\big(\sray(\bm{x})\big)$, $\tilde{\gt}(\bm{y})=\gt\big(\tray(\bm{y})\big)$ and $\tilde{T_2}(\bm{x},\bm{y})=T_2\big(\sray(\bm{x}),\tray(\bm{y})\big)$. The functions $J_{\sray}$ and $J_{\tray}$ denote the Jacobians of the coordinate transformations $\sray=\sray(\bm{x})$ and $\tray=\tray(\bm{y})$, respectively, as in Eq. (\ref{eq:stereo point}). For a point source, these coordinate transformations are the same, so $J_{\sray}=J_{\tray}$, with
\begin{equation}
  J_{\sray}(\bm{x}) = \left|\frac{\partial\sray}{\partial x_1}\times\frac{\partial\sray}{\partial x_2}\right| = \frac{4}{(1+|\bm{x}|^2)^2}.
\end{equation}
We assume that the Jacobian determinant $\det(\Diff\bm{m})$ is positive to obtain the Monge-Amp\`{e}re type equation
\begin{equation}\label{eq:MA point}
  \det\big(\Diff\bm{m}(\bm{x})\big) = T_1\,\tilde{T_2}\big(\bm{x},\bm{m}(\bm{x})\big)\frac{J_{\sray}(\bm{x})}{J_{\tray}\big(\bm{m}(\bm{x})\big)}\frac{\tilde{f}(\bm{x})}{\tilde{\gt}\big(\bm{m}(\bm{x})\big)} =: F_1\big(\bm{x},\bm{m}(\bm{x})\big).
\end{equation}
The approach is similar for a parallel source beam. The integrals on the left-hand side of Eq. (\ref{eq:energy int}) and in Eq. (\ref{eq:beta}) should be over an area element $\diff A(\bm{x})$ instead of a surface element $\diff S(\sray)$. We now only have the Jacobian for the change of variables on the right-hand side of Eq. (\ref{eq:energy int}), as defined in Eq. (\ref{eq:stereo par main}). Following the same calculation steps, we then arrive at a variant of Eq. (\ref{eq:MA point}) given by
\begin{equation}\label{eq:MA parallel}
  \det\big(\Diff\bm{m}(\bm{x})\big) = T_1\,\tilde{T_2}\big(\bm{m}(\bm{x})\big)\frac{1}{J_{\tray}\big(\bm{m}(\bm{x})\big)}\frac{f(\bm{x})}{\tilde{\gt}\big(\bm{m}(\bm{x})\big)} =: F_2\big(\bm{x},\bm{m}(\bm{x})\big),
\end{equation}
with $\tilde{T_2}\big(\bm{m}(\bm{x})\big) = T_2\Big(\tray\big(\bm{m}(\bm{x})\big)\Big)$. We will use $F$ to indicate either $F_1$ or $F_2$, depending on the optical system. When we combine Eq. (\ref{eq:MA point}) or Eq. (\ref{eq:MA parallel}) with Eq. (\ref{eq:Dm}), we obtain a condition on $\bm{P}$ given by
\begin{equation}
  \det\big(\bm{P}(\bm{x})\big)=F\big(\bm{x},\bm{m}(\bm{x})\big)\det\big(\bm{C}\big(\bm{x},\bm{m}(\bm{x})\big)\big).
\end{equation}
\indent To close the model, we have the boundary condition $\bm{m}\big(\partial\Xc\big)=\partial\Yc$, because of the edge-ray principle \cite{ries94}. This has the consequence that all transmitted light is mapped from the source to the target.

\section{The least-squares algorithm}\label{ls}
To recap the previous section, we derived a boundary value problem given by
\begin{subequations}\label{eq:system}
\begin{align}
  &\bm{C}\Diff\bm{m} = \bm{P}, \qquad \bm{x}\in\Xc, \label{eq:CDm2} \\
  &\text{subject to}\quad \det\bm{P} = F\det\bm{C}\quad\text{with}\quad\bm{P}\text{ SPD or SND}, \\
  &\bm{m}\big(\partial\Xc\big) = \partial\Yc. \label{eq:bnd2}
\end{align}
\end{subequations}
This BVP is solved to compute a mapping $\bm{m}$. Subsequently, substituting $\bm{y}=\bm{m}(\bm{x})$ in Eq. (\ref{eq:nabla}) we compute $u_1$, defining the surface shape. The least-squares method to solve a system like this has been explained in detail by Yadav \cite{yad18} and Romijn \cite{rom21} among others. This method has been used for cases where Fresnel reflections are not included. In Sec. \ref{sec:energy} we introduced changes in the Monge-Amp\`ere equation by considering these reflections. To take these changes into account, we slightly adapt the least-squares solver. In this section, we will give a brief overview of the method that we use to solve the boundary value problem (\ref{eq:system}). \\
\indent We introduce a functional $J_\text{I}$ as a measure of how closely we are approximating a solution to Eq. (\ref{eq:CDm2}). This functional is given by
\begin{equation}\label{eq:JI}
  J_{\text{I}}[\bm{m},\bm{P}] = \frac12 \iint_{\Xc}\|\bm{C}\Diff\bm{m}-\bm{P}\|_{\text{F}}^2\,\diff\bm{x},
\end{equation}
where $\|{\,.\,}\|_F$ is the Frobenius norm. As a measure of the difference between $\bm{m}\big(\partial\Xc\big)$ and $\partial\Yc$ we introduce the functional
\begin{equation}\label{eq:JB}
  J_\text{B}[\bm{m},\bm{b}] = \frac12\int_{\partial\Xc}|\bm{m}-\bm{b}|^2\,\diff s,
\end{equation}
where $\bm{b}:\partial\Xc\to\partial\Yc$ and $|{\,.\,}|$ indicates the standard 2-norm. We combine the two functionals by taking a weighted average with parameter $\alpha\in(0,1)$. The resulting functional is given by
\begin{equation}\label{eq:J}
  J[\bm{m},\bm{P},\bm{b}] = \alpha J_\text{I}[\bm{m},\bm{P}]+(1-\alpha)J_\text{B}[\bm{m},\bm{b}].
\end{equation}
We define the following function spaces for $\bm{P}$, $\bm{b}$ and $\bm{m}$:
\begin{subequations}
\begin{align}
    \mathcal{P}(\bm{m}) &= \left\{\bm{P}\in[C^1(\Xc)]^{2\times2} \mid \det(\bm{P})=F({\,\cdot\,},\bm{m})\det\big(\bm{C}({\,\cdot\,},\bm{m})\big),\ \bm{P}\text{ SPD or SND}\right\}, \\
    \mathcal{B} &= \left\{\bm{b}\in[C^1(\partial\Xc)]^2 \mid \bm{b}(\bm{x})\in\partial\Yc\right\}, \\
    \mathcal{M} &= [C^2(\Xc)]^2 .
\end{align}
\end{subequations}
In other words, $\mathcal{P}(\bm{m})$ contains $2\times2$ matrices with differentiable entries that satisfy the stated constraints. The set $\mathcal{B}$ contains vector-valued functions from the source boundary to the target boundary. Finally, $\mathcal{M}$ contains two times differentiable vector-valued functions defined on $\Xc$. \\
\indent We cover the domain $\Xc$ by a grid with gridpoints $\bm{x}_{ij}$. We calculate $T_1$ once, since it does not depend on $\bm{m}$. The algorithm then starts with an initial guess $\bm{m}^0$ and consequently $\bm{C}^0=\bm{C}({\,\cdot\,},\bm{m}^0)$. Also, we calculate $\tilde{T}_2({\,\cdot\,},\bm{m}^0)$ and $\beta^0=\beta(\bm{m}^0)$, which defines $\gt^0=\beta^0 g$ and subsequently a function $F^0=F({\,\cdot\,},\bm{m}^0)$. Then, for every $i=0,1,2,\dots$ we iterate
\begin{subequations}\label{eq:algorithm}
  \begin{align}
    \bm{P}^{i+1} &= \argmin_{\bm{P}\in\mathcal{P}(\bm{m}^i)}J_\text{I}[\bm{m}^i,\bm{P}], \label{eq:min P}\\
    \bm{b}^{i+1} &= \argmin_{\bm{b}\in\mathcal{B}}J_\text{B}[\bm{m}^i,\bm{b}], \label{eq:min b}\\
    \bm{m}^{i+1} &= \argmin_{\bm{m}\in\mathcal{M}}J[\bm{m},\bm{P}^{i+1},\bm{b}^{i+1}], \label{eq:min m}\\
    \bm{C}^{i+1} &= \bm{C}({\,\cdot\,},\bm{m}^{i+1}), \\
    \tilde{T}_2^{i+1} &= \tilde{T}_2({\,\cdot\,},\bm{m}^{i+1}), \quad \beta^{i+1} = \beta(\bm{m}^{i+1}), \quad F^{i+1} = F({\,\cdot\,},\bm{m}^{i+1}).
  \end{align}
\end{subequations}
In the first two steps, the functionals do not contain derivatives of the variables to be minimized and thus can be solved point-wise. The minimization for $\bm{m}^{i+1}$ does contain derivatives of $\bm{m}$, so this step cannot be done point-wise. The optimization step (\ref{eq:min m}) is done with calculus of variations \cite{rom21}. The first variation of $J[{\,\cdot\,},\bm{P},\bm{b}]$ with respect to an arbitrary $\bm{\eta}\in\mathcal{M}$ should be equal to 0. This condition gives the boundary value problem
\begin{subequations}
  \begin{align}
    &\diver(\bm{C}\tran\bm{C}\Diff\bm{m}) = \diver(\bm{C}\tran\bm{P}), &\quad&\text{for }\bm{x}\in\Xc, \label{eq:internal}\\
    &\alpha\bm{C}\tran\bm{C}(\Diff\bm{m})\bm{\hat{\nu}}+(1-\alpha)\bm{m}= \alpha\bm{C}\tran\bm{P}\bm{\hat{\nu}}+(1-\alpha)\bm{b}, &\quad&\text{for }\bm{x}\in\partial\Xc, \label{eq:minM boundary}
  \end{align}
\end{subequations}
with $\diver$ the divergence operator applied to the rows of a matrix and $\bm{\hat{\nu}}$ the outward unit normal of $\Xc$. We then discretize this system with finite volumes and solve the resulting linear system with a QR-decomposition. The last steps of the iteration scheme (\ref{eq:algorithm}) are straightforward evaluations from Eq. (\ref{eq:refl point}) or (\ref{eq:refl par}), Eq. (\ref{eq:beta}) and Eq. (\ref{eq:MA point}) or (\ref{eq:MA parallel}). \\
\indent After a given number of iterations or when a stopping criterion (i.e. a certain value of $J$) is met, we calculate $u$, defining the surface shape, from the resulting $\bm{m}$. We do this by solving Eq. (\ref{eq:nabla}) for $u_1$. An exact solution to this equation might not exist due to previous approximations. Therefore, we introduce a functional $I$ to minimize, given by
\begin{equation}\label{eq:calc u}
  I[\phi] = \frac12\iint_{\Xc}|\nabla_{\bm{x}}c({\,\cdot\,},\bm{m})-\nabla\phi|^2\,\diff\bm{x}.
\end{equation}
One can easily see that this functional becomes 0 when Eq. (\ref{eq:nabla}) is satisfied. Similar to the minimization for $\bm{m}$, we set the first variation of $I$ equal to 0 and use calculus of variations to derive a boundary value problem for $u_1$ of the form
\begin{subequations}\label{eq:bvp refl}
  \begin{align}
    &\Delta u_1 = \diver\big(\nabla_{\bm{x}}c({\,\cdot\,},\bm{m})\big),\qquad&&\bm{x}\in\Xc, \\
    &\nabla u_1\bm{\cdot}\bm{\hat{\nu}} = \nabla_{\bm{x}}c({\,\cdot\,},\bm{m})\bm{\cdot}\bm{\hat{\nu}}, \qquad&&\bm{x}\in\partial\Xc,
  \end{align}
\end{subequations}
as shown by Yadav \cite{yad18}. This is then solved for $u_1$, which is unique up to a constant. The constant can be chosen to fix the average distance of the freeform surface. From $u_1$ we can then calculate the function $u$, defining the freeform surface. For the point source we have $u=e^{u_1}$ and for the parallel source $u=u_1$.

\section{Numerical results}\label{results}
\begin{figure}[ht!]
  \centering
  \begin{subfigure}[t]{0.435\textwidth}
    \centering
    \includegraphics[width=\textwidth]{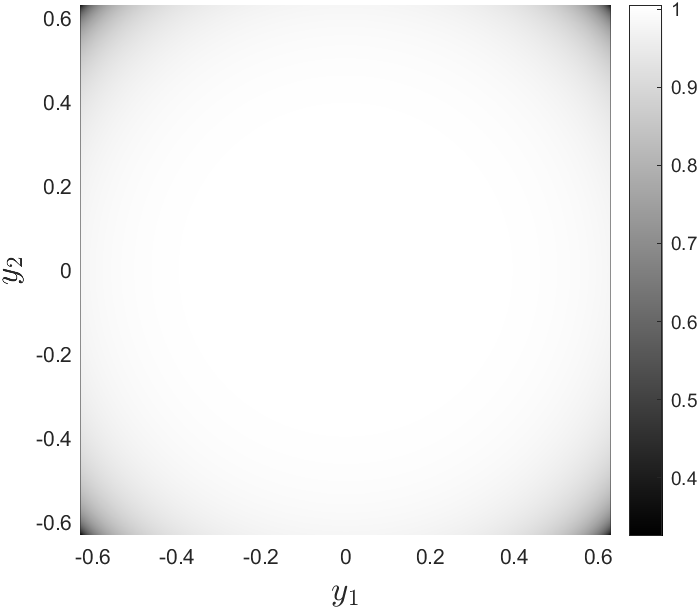}
    \caption{The hypothetical analytic target $J_{\tray} \tilde{g}$ corresponding to $\bm{m}(\bm{x})=\bm{x}$ with Fresnel reflections.}
    \label{fig:test intensity}
  \end{subfigure}
  \hfill
  \begin{subfigure}[t]{0.5\textwidth}
    \centering
    \includegraphics[width=\textwidth]{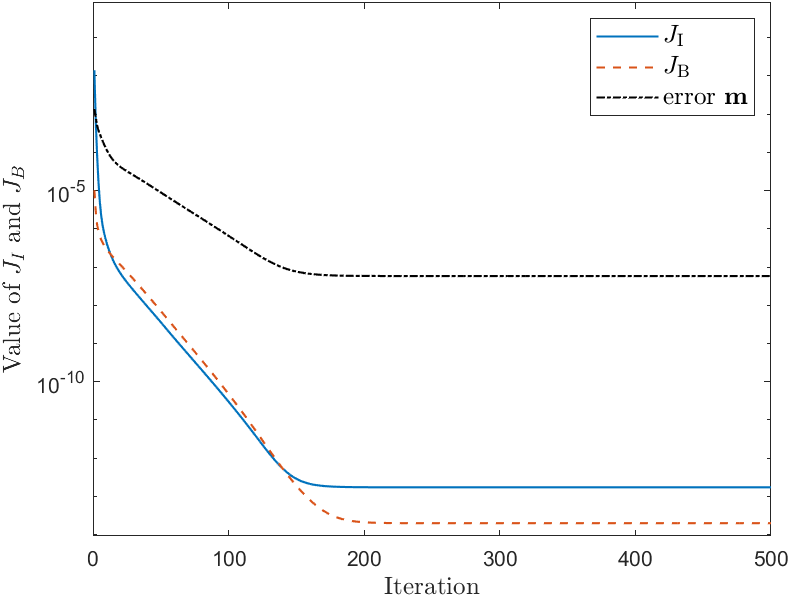}
    \caption{The values of $J_\text{I}$ and $J_\text{B}$ and the error compared to the exact mapping.}
    \label{fig:test conv}
  \end{subfigure}
  \caption{Target distribution and convergence of the analytic test case.}
\end{figure}

\begin{figure}[ht!]
  \centering
  \begin{subfigure}[t]{0.45\linewidth}
    \centering
    \includegraphics[width=\textwidth]{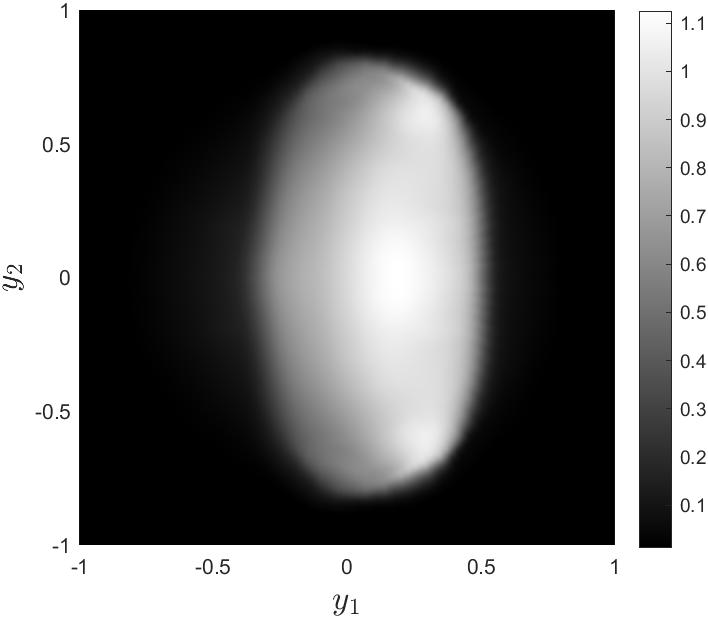}
    \caption{The hypothetical target $J_{\tray} \tilde{g}$ in stereographic coordinates.}
    \label{fig:road int ex}
  \end{subfigure}
  \hfill
  \begin{subfigure}[t]{0.45\linewidth}
    \centering
    \includegraphics[width=\textwidth]{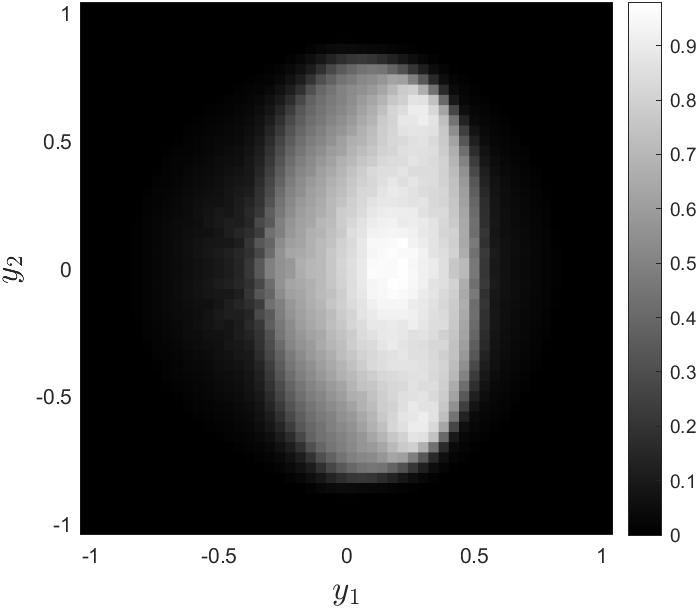}
    \caption{The raytraced target from a lens resulting from our algorithm.}
    \label{fig:road int ray}
  \end{subfigure}
  \caption{Target distribution of the streetlight example in stereographic coordinates. Note that the scales are different, but the shape is similar.}
\end{figure}

\begin{figure}[ht!]
  \centering
  \begin{subfigure}[t]{0.45\linewidth}
    \centering
    \includegraphics[width=\textwidth]{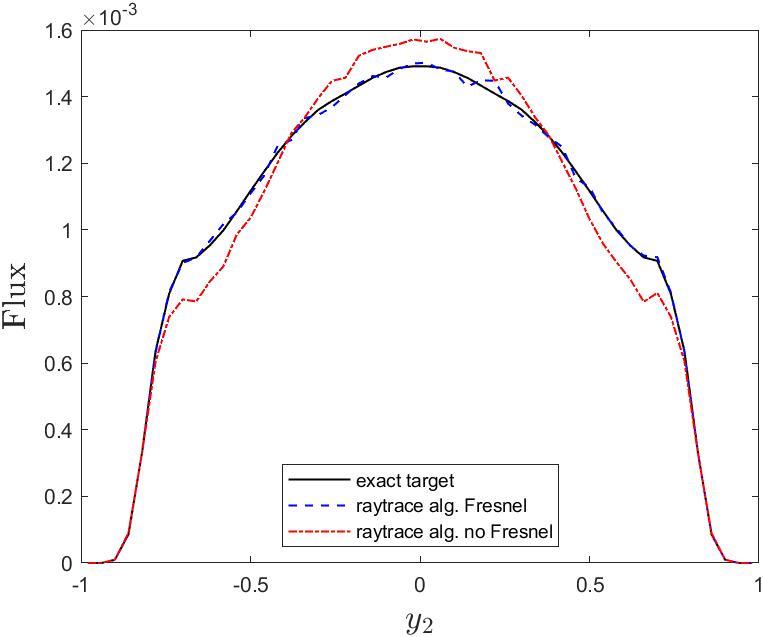}
    \caption{$y_1=0$.}
    \label{fig:comparison1}
  \end{subfigure}
  \hfill
  \begin{subfigure}[t]{0.45\linewidth}
    \centering
    \includegraphics[width=\textwidth]{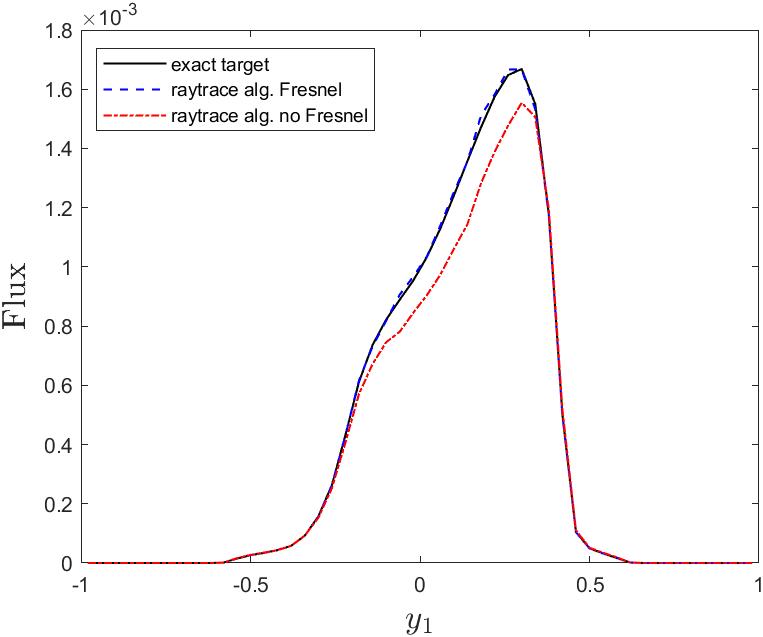}
    \caption{$y_2=-0.62$.}
    \label{fig:comparison2}
  \end{subfigure}
  \caption{A comparison between the desired output intensity and the results of the algorithm with and without Fresnel reflection scaled to have the same total flux. The results are obtained by raytracing with Fresnel reflection.}
  \label{fig:comparison}
\end{figure}

\begin{figure}[ht!]
\centering
\begin{minipage}{0.47\linewidth}
  \centering
  \includegraphics[width=\linewidth]{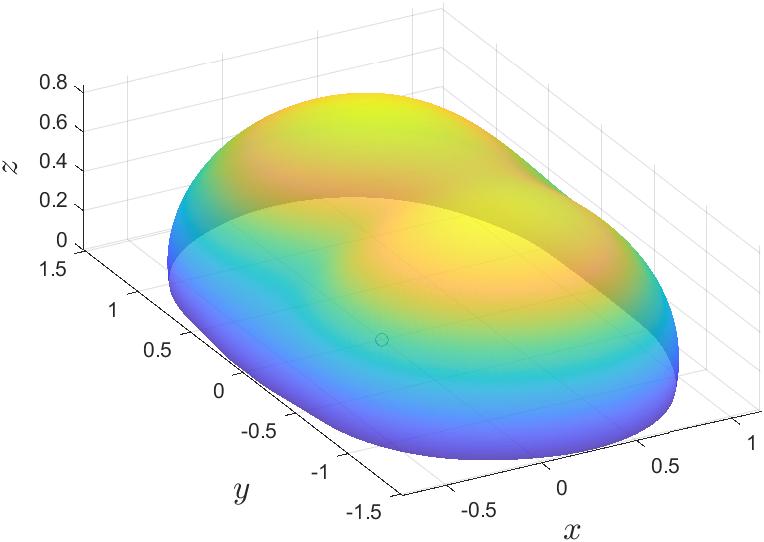}
  \caption{The freeform lens surface resulting from our algorithm in Cartesian coordinates.}
  \label{fig:peanut}
\end{minipage}
\hfill
\begin{minipage}{0.47\textwidth}
  \centering
  \includegraphics[width=\textwidth]{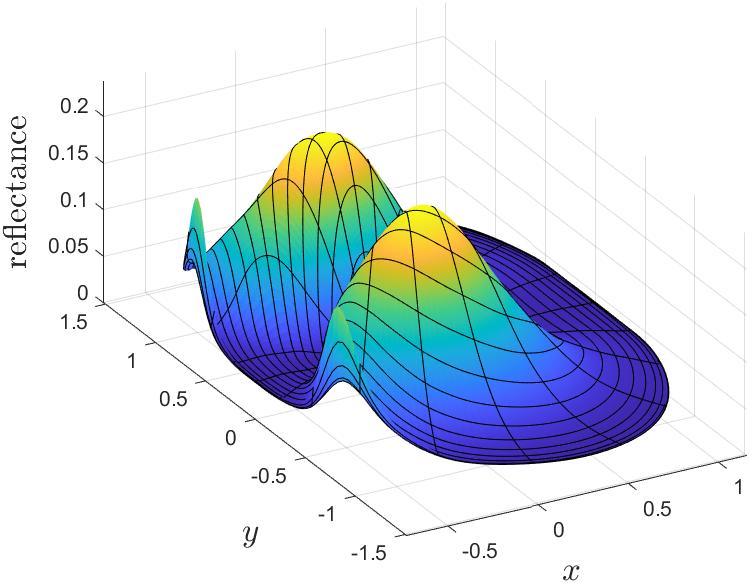}
  \caption{The reflectance on the freeform surface resulting from our algorithm as a function of Cartesian $x,y$-coordinates.}
  \label{fig:reflectance}
\end{minipage}
\end{figure}

We apply our algorithm to two different cases. First, we use a parallel source for which we can compare the result to an analytic solution. Then, we use the algorithm to construct a lens for street lighting and compare the result to previous results without Fresnel reflections. \\
\indent First, we apply our algorithm to a problem for which we can find an analytic solution. Without Fresnel reflections, choosing the mapping $\bm{m}(\bm{x})=\bm{x}$ for a parallel source corresponds to a source distribution $f$ and stereographic target distribution $J_{\tray}\tilde{g}$ that are equal, in accordance with Eq. (\ref{eq:MA parallel}). This is no longer the case when we consider Fresnel reflections. The rays close to $\tray=(0,0,1)\tran$ have smaller incident and transmitted angles than rays closer to the edge of the domain and as a result the reflection coefficient is smaller near $\tray=(0,0,1)\tran$. If we would have a uniform source $f(\bm{x})=f_0$, then the transmitted stereographic target distribution is no longer uniform, but given by $J_{\tray}(\bm{x})\tilde{\gt}(\bm{x})=T_1\,\tilde{T}_2(\bm{x})f_0$. From substituting $\bm{m}(\bm{x})=\bm{x}$ into Eq. (\ref{eq:stereo par inv}) we obtain
\begin{equation}
  t_3 = n-H(|\bm{x}|;n).
\end{equation}
Using this, we derive an analytic expression for the transmitted target distribution, see Fig. \ref{fig:test intensity}.
 \\
\indent We choose the source and stereographic target domain to be the square $[-0.63,0.63]^2$. The source domain is covered with a uniform $200\times200$ grid. The lens has a refractive index $n=1.5$. Our initial guess for the mapping is a random perturbation of $\bm{m}(\bm{x})=\bm{x}$. Every target gridpoint $\bm{y}_{ij}=\bm{m}(\bm{x}_{ij})$ is shifted from $\bm{y}=\bm{x}_{ij}$ in both the $y_1$- and $y_2$-direction with a maximum of half the grid size. The convergence behavior of the algorithm is shown in Fig. \ref{fig:test conv} by plotting the error of the mapping $\bm{m}^i$ compared to the exact mapping $\bm{m}(\bm{x})=\bm{x}$. The figure also shows the values of the functionals $J_\text{I}$ and $J_\text{B}$. We can see that the error converges to approximately $10^{-7}$, while both functionals converge to values around $10^{-14}$. This makes sense, since both functionals consist of the square of an error term of the mapping. \\
\indent Next, we compute a lens for a typical street lamp. We model an LED as a point source with a Gaussian light distribution on the positive half-sphere, with variance 0.05 and scaled to a total flux of 1. The target intensity is an intensity that is used for street lights \cite{rom19}, see Fig. \ref{fig:road int ex}. Again, the lens material has refractive index $n=1.5$. We use a uniform $200\times200$ grid and run the algorithm for 300 iterations.
\\
\indent The resulting lens has the shape of a so-called `peanut lens', see Fig. \ref{fig:peanut}. We verify this result with our own raytracer in MATLAB, tracing ten million rays in a quasi-Monte Carlo sense, including Fresnel reflections. The reflected flux is 13.4\% of the source flux. The resulting target intensity in stereographic coordinates is shown in Fig. \ref{fig:road int ray}. We can see that it corresponds well with the desired target intensity in Fig. \ref{fig:road int ex}, although with a lower intensity. \\
\indent For comparison, we also used the least-squares algorithm from \cite{rom19} without taking into account Fresnel reflections. All other input parameters were chosen to be the same. Then, a raytrace with Fresnel reflection was applied to the resulting lens surface, again tracing ten million rays. The flux lost due to reflection is 12.6\% of the source flux here. Evaluating the result, we see that the intensity pattern deviates significantly from the desired one along some intersection planes, as shown in Fig. \ref{fig:comparison}. In that figure, the intensities are scaled to have the same total flux. This way, we can compare the intensity shapes without considering the flux. In comparison to the result of the algorithm without Fresnel, the results from the algorithm as stated in this article seem to match the desired intensity better. Therefore, this example shows the usefulness of the adaptations to the algorithm introduced in this paper.
\\
\indent The reflectance on the resulting freeform surface is shown in Fig. \ref{fig:reflectance}. We see that it varies a lot over the surface, with maximal values around $0.2$. This indicates that without taking Fresnel reflections into account, the transmitted intensity would be significantly higher than desired. By incorporating these reflections into our algorithm we make sure that the output is of the correct shape. Therefore, this figure shows the importance of the modifications we have elaborated in this paper.
\captionsetup{font=small,margin=10pt,labelsep=period}
\captionsetup{font=small,width=4.25in,labelsep=period}

\section{Conclusion}
In this paper we presented a method to take into account Fresnel reflections when designing a freeform lens. We have derived an expression for the reflectance in terms of source and target coordinates. This has been incorporated in the least-squares algorithm that has been used before as an inverse method for designing freeform optical surfaces. \\
\indent We tested this modified algorithm on two cases. For a parallel source beam we constructed a test where we know the analytic solution. This was used to verify our algorithm. We then investigated a practical application, namely that of street lighting. It was shown that the reflectance on parts of the lens can be very significant. This shows the importance of including Fresnel reflections in our algorithm to ensure the correct target intensity shape. \\
\indent In this paper we limited ourselves to energy loss due to Fresnel reflections, but the same techniques could be used to take into account other phenomena like partially absorbing lenses. Likewise, we only elaborated the algorithm for lenses with a far-field target and point source or collimated input beam, but similar algorithms can be used to consider Fresnel reflections in the design of other optical systems. For future research, it would also be interesting to investigate the possibility of minimizing the reflectance and with that the loss of light.

\section*{Disclosures}
The authors declare no conflicts of interest.

\bibliographystyle{unsrt}
\bibliography{../../../MyLateX/bibliographies/full}

\end{document}